\documentclass[fleqn,11pt]{article}

\usepackage{latexsym,ifthen,epsfig,color}
\usepackage{amsmath,amssymb,amsthm}
\usepackage{float}

\newcommand{\NP}{\ensuremath{\mathbb{NP}}}

\newtheorem{theorem}{Theorem}
\newtheorem{lemma}{Lemma}
\newtheorem{corollary}{Corollary}

\newtheorem{clai}{Claim}
\newtheorem{observation}{Observation}

\begin{document}

\title{Efficient Domination for Some Subclasses of $P_6$-Free Graphs in Polynomial Time}

\author{
Andreas Brandst\"adt\footnote{Fachbereich Informatik, 
Universit\"at Rostock, A.-Einstein-Str. 22, D-18051 Rostock, Germany.
\texttt{e-mail: ab@informatik.uni-rostock.de}
}
\and
Elaine M. Eschen\footnote{MIS, West Virginia University, Morgantown, WV, USA.
\texttt{e-mail: eschen@mail.wvu.edu}
}
\and
Erik Friese\footnote{Fachbereich Mathematik, 
Universit\"at Rostock, Ulmenstr. 69, D-18057 Rostock, Germany.
\texttt{e-mail: erik.friese@uni-rostock.de}
}
}

\maketitle













\begin{abstract}
Let $G$ be a finite undirected graph. A vertex {\em dominates} itself and all its neighbors in $G$.
A vertex set $D$ is an {\em efficient dominating set} (\emph{e.d.}\ for short) of $G$ if every vertex of $G$ is dominated by exactly one vertex of $D$.
The \emph{Efficient Domination} (ED) problem, which asks for the existence of an e.d.\ in $G$, is known to be \NP-complete even for very restricted graph classes such as $P_7$-free chordal graphs. The ED problem on a graph $G$ can be reduced to the Maximum Weight Independent Set (MWIS) problem on the square of $G$.
The complexity of the ED problem is an open question for $P_6$-free graphs and was open even for the subclass of $P_6$-free chordal graphs. In this paper, we show that squares of $P_6$-free chordal graphs that have an e.d. are chordal; this even holds for the larger class of ($P_6$, house, hole, domino)-free graphs.
This implies that ED/WeightedED is solvable in polynomial time for ($P_6$, house, hole, domino)-free graphs; in particular, for $P_6$-free chordal graphs.
Moreover, based on our result that squares of $P_6$-free graphs that have an e.d. are hole-free and some properties concerning odd antiholes, we show that
squares of ($P_6$, house)-free graphs (($P_6$, bull)-free graphs, respectively) that have an e.d. are perfect.
This implies that ED/WeightedED is solvable in polynomial time for ($P_6$, house)-free graphs and for ($P_6$, bull)-free graphs (the time bound for ($P_6$, house, hole, domino)-free graphs is better than that for ($P_6$, house)-free graphs).
The complexity of the ED problem for $P_6$-free graphs remains an open question.
\end{abstract}


\section{Introduction}\label{sec:intro}

Let $G=(V,E)$ be a finite undirected graph. A vertex $v \in V$ {\em dominates} itself and its neighbors. A vertex subset $D \subseteq V$ is an {\em efficient dominating set} ({\em e.d.} for short) of $G$ if every vertex of $G$ is dominated by exactly one vertex in $D$.
Note that not every graph has an e.d.; the {\sc Efficient Dominating Set} (ED) problem asks for the existence of an e.d.\ in a given graph $G$.
If a vertex weight function $\omega: V \to \mathbb{N}$ is given, the {\sc Weighted Efficient Dominating Set} (WED) problem asks for a minimum weight e.d. in $G$, if there is one, or for determining that $G$ has no e.d. The importance of the ED problem mostly results from the fact that the ED problem for a graph $G$ is a special case of the {\sc Exact Cover} problem for hypergraphs (problem [SP2] of \cite{GarJoh1979}); ED is the Exact Cover problem for the closed neighborhood hypergraph of $G$.

For a graph $F$, a graph $G$ is called {\em $F$-free} if $G$ contains no induced subgraph isomorphic to $F$.

We denote by $G+H$ the disjoint union of graphs $G$ and $H$.
Let $P_k$ denote a chordless path with $k$ vertices, and let $2P_k$ denote $P_k+P_k$, and correspondingly for $kP_2$.
The {\em claw} is the $4$-vertex tree with three vertices of degree $1$.

Many papers have studied the complexity of ED on special graph classes - see e.g. \cite{BraMilNev2013,Milan2012} for references. In particular, a standard reduction from the Exact Cover problem shows that ED remains \NP-complete for $2P_3$-free chordal graphs and for bipartite graphs. Moreover, it is known to be \NP-complete for line graphs and thus, for claw-free graphs.

A {\it linear forest} is a graph whose components are paths; equivalently, it is a graph that is cycle-free and claw-free.
The \NP-completeness of ED on chordal graphs, on bipartite graphs and on claw-free graphs implies: If $F$ is not a linear forest, then ED is \NP-complete on $F$-free graphs.
This motivates the analysis of ED/WED on $F$-free graphs for linear forests $F$. For $F$-free graphs, where $F$  is a linear forest, the only remaining open case is the complexity of ED on $P_6$-free graphs (see \cite{BraGia2014}).

\medskip

\noindent
The main results of this paper are the following:
\begin{enumerate}
\item[-] If $G$ is ($P_6$, HHD)-free and has an e.d., then $G^2$ is chordal. Then using a subsequently described reduction of ED/WED on $G$ to the Maximum Weight Independent Set (MWIS) problem on $G^2$, we obtain a polynomial time solution for ED/WED on this class of graphs, since MWIS is solvable in polynomial time on chordal graphs. This also gives a dichotomy result for $P_k$-free chordal graphs, since ED is \NP-complete for $P_7$-free chordal graphs.
\item[-] If $G$ is $P_6$-free and has an e.d., then $G^2$ is hole-free. This does not yet imply that ED for $P_6$-free graphs is solvable in polynomial time, since the MWIS problem for hole-free graphs is an open question but it leads to further results on ED for subclasses of $P_6$-free graphs.
\item[-] If $G$ is $P_6$-free and has an e.d., then odd antiholes in $G^2$ have very special structure. Analyzing the structure of $C_4$ realizations in $G^2$, we obtain a polynomial time solution of ED/WED for ($P_6$, house)-free graphs and for ($P_6$, bull)-free graphs, since in this case, $G^2$ is perfect if $G$ has an e.d.
\end{enumerate}


\section{Basic notions and results}\label{sec:basicnotions}

\subsection{Some basic notions}

All graphs considered in this paper are finite, undirected and simple (i.e., without loops and multiple edges). For a graph $G$, let $V(G)$ or simply $V$ denote its vertex set and $E(G)$ or simply $E$ its edge set; throughout this paper, let $|V|=n$ and $|E|=m$. We can assume that $G$ is connected (otherwise, ED can be solved separately for its components); thus, $m \ge n-1$. For $U \subseteq V$, let $G[U]$ denote the subgraph of $G$ induced by $U$.   

For a vertex $v \in V$, $N(v)=\{u \in V \mid uv \in E\}$ denotes its ({\em open}) {\em neighborhood}, and $N[v]=\{v\} \cup N(v)$ denotes its {\em closed neighborhood}. A vertex $v$ {\em sees} the vertices in $N(v)$ and {\em misses} all the others. Let $d_G(v,w)$ denote the distance between $v$ and $w$ in $G$.

Let $P_k$ denote a chordless path with $k$ vertices, and let $C_k$ denote a chordless cycle with $k$ vertices. Chordless cycles $C_k$ with $k \ge 5$ are called {\em holes}. The complement graph $\overline{P_5}$ is also called {\em house}. {\em Domino} has six vertices and can be obtained by adding a vertex $y$ to a $P_5$ $x_1,\ldots,x_5$ with edges $x_ix_{i+1}$, $1\le i \le 4$, such that $yx_1 \in E$, $yx_3 \in E$, and $yx_5 \in E$. A graph is {\em chordal} if it is $C_k$-free for every $k \ge 4$. A graph is ({\em house, hole, domino})-free ({\em HHD}-free for short) if it has no induced subgraph isomorphic to a house, hole or domino. Obviously, chordal graphs are HHD-free, and $G$ is ($P_6$, HHD)-free if and only if $G$ is ($P_6, C_5, C_6$, house, domino)-free. The importance of HHD-free graphs as a natural generalization of chordal graphs is illustrated by various characterizations of them such as:
$G$ is HHD-free if and only if $G$ is $(5,2)$-chordal (see e.g. \cite{BraLeSpi1999}).

\subsection{Reducing the ED problem on a graph to the MWIS problem on its square}\label{EDGMWISG2}

The {\em square} of a graph $G = (V,E)$ is the graph $G^2 = (V,E^2)$ such that $uv\in E^2$ if and only if $d_G(u,v)\in\{1,2\}$.
In \cite{BraLeiRau2012,Leite2012,Milan2012}, the following relationship between the ED problem on a graph $G$ and the maximum weight independent set (MWIS) problem on $G^2$ is used:

\begin{lemma}\label{mainequived}
Let $G=(V,E)$ be a graph and $\omega(v):= |N[v]|$ a vertex weight function for $G$. Then the following are equivalent for any subset $D \subseteq V$:
\begin{enumerate}
\item[$(i)$] $D$ is an efficient dominating set in $G$.
\item[$(ii)$] $D$ is a minimum weight dominating set in $G$ with $\omega(D)=|V|$.
\item[$(iii)$] $D$ is a maximum weight independent set in $G^2$ with $\omega(D)=|V|$.
\end{enumerate}
\end{lemma}

Thus, the ED problem on a graph class $\cal C$ can be reduced to the MWIS problem on the squares of graphs in
$\cal C$. In \cite{BraFicLeiMil2013}, this is extended to the vertex-weighted version WED of the ED problem.

\section{Squares of ($P_6$, HHD)-free graphs that have an e.d. are chordal}\label{SquaresP6frHHDfrchordal}

Obviously, the square of a chordal graph can contain a $C_4$ as for example, the complete 4-sun shows. If we additionally require that the graph is $P_6$-free and has an e.d., the situation is different: The main result of this section is Theorem~\ref{P6HHDfrG2chordal}, which shows that for any graph $G$ that is ($P_6$, HHD)-free and has an e.d., its square $G^2$ is chordal, i.e., $C_k$-free for every $k \ge 4$. Theorem~\ref{P6frG2holefree} in Section~\ref{sec:G2holefree} shows that $G^2$ is $C_k$-free for every $k \ge 5$ for the larger class of $P_6$-free graphs, but its proof is long and technically involved. For the special case of ($P_6$, HHD)-free graphs, we give a direct proof here since it is much shorter than the proof of Theorem~\ref{P6frG2holefree}.
\begin{theorem}\label{P6HHDfrG2chordal}
If $G$ is a $(P_6$, HHD$)$-free graph that has an e.d., then $G^2$ is chordal.
\end{theorem}

For the proof of Theorem~\ref{P6HHDfrG2chordal}, we first prove several lemmas. Let $G$ be a ($P_6$, HHD)-free graph with an e.d. $D$.
Suppose that $G^2$ contains a chordless cycle $C_k$ $C$ with vertices $v_1,\ldots,v_k$, $k \ge 4$; we call these the {\em real vertices} of $C$ and denote them by $R(C)=\{v_1,\ldots,v_k\}$. For $d_G(v_i,v_{i+1})=2$ (index arithmetic is modulo $k$ throughout this section), let $x_i$ be a common neighbor of $v_i$ and $v_{i+1}$; we call these $x_i$ vertices the {\em auxiliary vertices} of $C$ and denote the set of these vertices by $A(C)$. Let $V(C)=R(C) \cup A(C)$ denote the set of vertices (real and auxiliary) in $G$ {\em realizing} a $C_k$ $C$ in $G^2$; we call $V(C)$ a {\em cycle embedding}.
\begin{observation}\label{easyfacts}
For every $i \in \{1,\ldots,k\}$, $d_G(v_i,v_{i+1}) \le 2$. Also, $d_G(v_i,v_j) > 2$ if $v_i$ and $v_j$ are not consecutive in the $C_k$ $C$ in $G^2$. In particular, if $d_G(v_i,v_{i+1})=1$ then $d_G(v_{i+1},v_{i+2})=2$ and $d_G(v_{i-1},v_i)=2$.  Clearly, auxiliary vertices are pairwise distinct, and for every $x_i$, 
$v_jx_i \notin E$ for all $j \notin \{i,i+1\}$.
\end{observation}

We claim that there are $k$ distinct auxiliary vertices $x_1,\ldots,x_k$ in $V(C)$:
\begin{lemma}\label{Ckvivi+1dist2}
For all $i \in \{1,\ldots,k\}$, $d_G(v_i,v_{i+1}) = 2$.
\end{lemma}

\noindent
{\bf Proof.}
Without loss of generality, suppose $v_1v_2 \in E$. Then $d_G(v_2,v_3)=2$ and $d_G(v_k,v_1)=2$.

\medskip

\noindent
{\bf Case $k=4$:} If further $v_3v_4 \in E$, then $V(C)$ induces either a $C_6$ or a domino in $G$ which is a contradiction. Thus, $d_G(v_3,v_4)=2$ and there is a vertex $x_3$.

\noindent
Since $\{v_1,v_2,x_2,v_3,x_3,v_4\}$ does not induce a $P_6$, we have $x_2x_3 \in E$.

\noindent
Since $\{v_2,v_1,x_4,v_4,x_3,v_3\}$ does not induce a $P_6$, we have $x_3x_4 \in E$.

\noindent
Since $\{v_3,x_2,v_2,v_1,x_4,v_4\}$ does not induce a $P_6$, we have $x_2x_4 \in E$.

Now, $\{v_1,v_2,x_2,x_3,x_4\}$ induces a house which is a contradiction. Thus, we have $d_G(v_i,v_{i+1})=2$ for all $i \in \{1,\ldots,4\}$ and Lemma~\ref{Ckvivi+1dist2} holds for $k=4$.

\medskip

For $k \ge 5$, since $\{v_k$, $x_k$, $v_1$, $v_2$, $x_2$, $v_3\}$ does not induce a $P_6$, we have $x_kx_2 \in E$.
If $x_k$ and $x_2$ have a common neighbor $x_i$, $2<i<k$, then $\{v_1,v_2,x_2,x_k,x_i\}$ induces a house; thus for all $i$ with $2<i<k$, we have:
$$(*) \mbox{ } x_2x_i \notin E \mbox{ or } x_kx_i \notin E.$$

\medskip
\noindent
{\bf Case $k=5$:} Without loss of generality, suppose $d_G(v_3,v_4)=2$. 

\noindent
Since $\{v_1,v_2,x_2$, $v_3,x_3,v_4\}$ does not induce a $P_6$, we have $x_2x_3 \in E$; thus, $x_3x_5 \notin E$ by $(*)$. 

\noindent
Since $\{v_1,v_2,x_2,x_3,v_4,v_5\}$ does not induce a $P_6$, we have $d_G(v_4,v_5)=2$ and thus, there is a vertex $x_4$. 

\noindent
Since $\{v_2,v_1,x_5,v_5,x_4,v_4\}$ does not induce a $P_6$, we have $x_4x_5 \in E$, which implies $x_2x_4 \notin E$ by $(*)$. 

\noindent
Since $\{x_2,x_3,v_4,x_4,x_5\}$ does not induce a $C_5$, we have $x_3x_4 \in E$; but now, $G[\{x_2,x_3,v_4,x_4,x_5\}]$ is a house, which is a contradiction. Thus, Lemma~\ref{Ckvivi+1dist2} holds for $k=5$.

\medskip
\noindent
{\bf Case $k>5$:} Since $\{v_4,v_3,x_2$, $x_k,v_k,v_{k-1}\}$ does not induce a $P_6$, we have either $d_G(v_3,v_4)=2$ or $d_G(v_k,v_{k-1})=2$; without loss of generality, let $d_G(v_3,v_4)=2$ and there is a vertex $x_3$. 

\noindent
Since $\{v_4,x_3,v_3,x_2,v_2,v_1\}$ does not induce a $P_6$, we have $x_2x_3 \in E$, which implies $x_3x_k \notin E$ by $(*)$.

\noindent
Since $\{v_4,x_3,x_2,x_k,v_k,v_{k-1}\}$ does not induce a $P_6$, we have $d_G(v_k,v_{k-1})=2$ and thus, there is a vertex $x_{k-1}$.

\noindent
Since $\{v_{k-1},x_{k-1},v_k,x_k,v_1,v_2\}$ does not induce a $P_6$, we have $x_{k-1}x_k \in E$, which implies $x_2x_{k-1} \notin E$ by $(*)$.

\noindent
Since $\{v_4,x_3,x_2$, $x_k,x_{k-1},v_{k-1}\}$ does not induce a $C_6$ or $P_6$, we have $x_3x_{k-1} \in E$ and now $G[\{v_1,v_2,x_2,x_k,x_3,x_{k-1}\}]$ is a domino, which is a contradiction.  Thus, Lemma~\ref{Ckvivi+1dist2} holds for $k>5$.
\qed

\begin{lemma}\label{Ckxixi+1dist1}
For all $i \in \{1,\ldots,k\}$, $x_ix_{i+1} \in E$.
\end{lemma}

\noindent
{\bf Proof.}
Without loss of generality, suppose to the contrary that $x_1x_2 \notin E$. Then, since $\{v_1,x_1,v_2,x_2,v_3,x_3\}$ does not induce a $P_6$, we have $x_1x_3 \in E$ or $x_2x_3 \in E$.  If $x_1x_3 \in E$ then $x_2x_3 \in E$, else $\{x_1,v_2,x_2, v_3,x_3\}$ induces a $C_5$.  Thus, $x_2x_3 \in E$. 
Since $\{v_1,x_1,v_2,x_2,x_3,v_4\}$ does not induce a $P_6$, we have $x_1x_3 \in E$. But, now $\{x_1,v_2,x_2, v_3,x_3\}$ induces a house, which is a contradiction; thus, $x_1x_2 \in E$ and Lemma \ref{Ckxixi+1dist1} is shown. 
\qed

\begin{lemma}\label{noDvertexinC}
$D \cap \{v_1,\ldots,v_k,x_1,\ldots,x_k\} = \emptyset$.
\end{lemma}

\noindent
{\bf Proof.}
First suppose to the contrary that $D \cap \{v_1,\ldots,v_k\} \neq \emptyset$; without loss of generality, let $v_1 \in D$. Then $v_2 \notin D$ and $v_k \notin D$, but they must be dominated by $D$-vertices, say $d_2, d_k \in D$ with $d_2v_2 \in E$ and $d_kv_k \in E$. Since $d_G(v_2,v_k) > 2$, $d_2 \neq d_k$. Also, $d_2 \notin \{x_1,x_2\}$ and $d_k \notin \{x_{k-1}, x_{k}\}$. Now $G[\{d_2,v_2,x_1,x_k,v_k,d_k\}]$ is a $P_6$, which is a contradiction. Thus, $D \cap \{v_1,\ldots,v_k\} = \emptyset$.


Now suppose to the contrary that $D \cap \{x_1,\ldots,x_k\} \neq \emptyset$; without loss of generality, let $x_1 \in D$. We know already that $v_3,v_k \notin D$ and thus, there is $d_3 \in D$ with $d_3v_3 \in E$ such that $d_3 \notin \{x_2,x_3\}$.
If $k = 4$ and $d_3v_4 \in E$, then $G[\{v_3,d_3,v_4,x_4,x_1,v_2\}]$ is a $P_6$; thus, $d_3v_4 \notin E$.
If $k \ge 5$ then $d_3v_k \notin E$ since $d_G(v_3,v_k) > 2$.
So there must be $d_k \in D$ with $d_kv_k \in E$ such that $d_k \neq d_3$ and $d_k \notin \{x_{k-1},x_k\}$.
If $x_2x_k \notin E$ then $G[\{d_3,v_3,x_2,x_1,x_k,v_k\}]$ is a $P_6$, and if $x_2x_k \in E$ then $G[\{d_3,v_3,x_2,x_k,v_k,d_k\}]$ is a $P_6$, which is a contradiction. Thus, Lemma \ref{noDvertexinC} is shown.
\qed

\medskip

For all $i \in \{1,\ldots,k\}$, let $d_i \in D$ be the vertex with $d_iv_i \in E$. We claim that $d_1,\ldots,d_k$ are pairwise distinct:
\begin{corollary}\label{personalD}
For all $i \in \{1,\ldots,k\}$, $d_i$ has exactly one neighbor in $\{v_1,\ldots,v_k\}$.
\end{corollary}

\noindent
{\bf Proof.}
As in Observation~\ref{easyfacts}, a $D$-vertex cannot see both $v_i$ and $v_j$
if $v_i$ and $v_j$ are not consecutive in the $C_k$ $C$ in $G^2$.
Suppose without loss of generality $d_1=d_2$, i.e., $d_1v_1 \in E$ and $d_1v_2 \in E$.
Then $x_1$ can be replaced by $d_1$ in the cycle embedding and thus, $d_1$ is an auxiliary vertex in $D$ which contradicts Lemma~\ref{noDvertexinC}.
\qed

\begin{lemma}\label{dimissesxi}
For all $i \in \{1,\ldots,k\}$, $d_ix_{i-1} \notin E$ or $d_ix_i \notin E$.
\end{lemma}

\noindent
{\bf Proof.}
Without loss of generality, assume that $d_1x_k \in E$ and $d_1x_1 \in E$. Then $G[\{d_2,v_2,x_1$, $x_k,v_k,d_k\}]$ is a $P_6$, which is a contradiction.
\qed

\begin{lemma}\label{notd1x1impliesnotd1x2}
For all $i \in \{1,\ldots,k\}$, if $d_ix_i \notin E$ then $d_ix_{i+1} \notin E$, and if $d_ix_{i-1} \notin E$ then $d_ix_{i-2} \notin E$.
\end{lemma}

\noindent
{\bf Proof.}
Assume that $d_ix_i \notin E$ and $d_ix_{i+1} \in E$. Then $\{d_i,v_i,x_i,x_{i+1},v_{i+1}\}$ induces a house, which is a contradiction. Thus, if $d_ix_i \notin E$ then $d_ix_{i+1} \notin E$ and similarly, if $d_ix_{i-1} \notin E$ then $d_ix_{i-2} \notin E$.
\qed

\begin{lemma}\label{diimpliesdi+2}
For all $i \in \{1,\ldots,k\}$, $d_ix_i \notin E$ implies $d_{i+2}x_{i+1} \in E$ and $d_{i+2}x_{i+2} \notin E$.
\end{lemma}

\noindent
{\bf Proof.}
Assume without loss of generality that $d_1x_1 \notin E$.
Then, by Lemma~\ref{notd1x1impliesnotd1x2}, $d_1x_2 \notin E$ and if $d_3x_1 \in E$ then $d_3x_2 \in E$.
Since $\{d_1,v_1,x_1,x_2,v_3,d_3\}$ does not induce a $P_6$, we have $d_3x_2 \in E$ or $d_3x_1 \in E$, which implies $d_3x_2 \in E$. Then, by Lemma~\ref{dimissesxi}, $d_3x_3 \notin E$.
\qed

\medskip

For an odd hole, repeating the argument of Lemma \ref{diimpliesdi+2} on $d_3x_3 \notin E$, and so on, determines all the edges and non-edges between $D$-vertices and auxiliary vertices. For $C_4$ or an even hole, repeating the argument determines the edges for every second $D$-vertex, but then a second round (using the fact that $d_2x_2 \notin E$) determines the remaining edges and non-edges.

\medskip

\noindent
{\bf Proof of Theorem~\ref{P6HHDfrG2chordal}.}
First suppose that $C$ is a $C_4$ in $G^2$. Then, since $G$ is HHD-free, we have $x_1x_3 \in E$ or $x_2x_4 \in E$; without loss of generality say $x_1x_3 \in E$. Moreover, $d_1x_1 \notin E$ or $d_2x_1 \notin E$; without loss of generality say $d_1x_1 \notin E$.
Then by Lemma~\ref{diimpliesdi+2}, $d_3x_2 \in E$ (and thus, $d_2x_2 \notin E$) and $d_3x_3 \notin E$ holds and repeating the same arguments, we get $d_4x_3 \in E$,
$d_4x_4 \notin E$, and $d_2x_1 \in E$, but now $G[\{d_1,v_1,x_1,x_3,v_3,d_3\}]$ is a $P_6$ which is a contradiction.

\medskip

Now suppose that $C$ is a $C_k$ in $G^2$ for some $k \ge 5$. Then, since $G$ is HHD-free, there is an edge $x_ix_j \in E$ where $j \notin \{i-1, i+1\}$.
Then in the case that $d_ix_i \notin E$ (and thus also $d_jx_j \notin E$), $G[\{d_i,v_i,x_i,x_j,v_j,d_j\}]$ is a $P_6$. The case when $d_{i+1}x_i \notin E$ is symmetric.  Thus, we have a contradiction. This concludes the proof of Theorem~\ref{P6HHDfrG2chordal}.
\qed

\begin{corollary}\label{EDP6HHDfrpol}
For $(P_6$, HHD$)$-free graphs, the WED problem is solvable in polynomial time.
\end{corollary}

\noindent
{\bf Proof.}
By Lemma \ref{mainequived}, the ED problem for $G$ can be reduced to the MWIS problem for $G^2$.
By Theorem~\ref{P6HHDfrG2chordal}, $G^2$ is chordal. By the result of Frank \cite{Frank1975}, the MWIS problem can be solved in linear time for chordal graphs.
Thus, for ($P_6$, HHD)-free graphs, the ED problem is solvable in polynomial time. By \cite{BraFicLeiMil2013}, the WED problem can be solved in polynomial time for the same class.
\qed

\section{Some properties of $P_6$-free graphs that have an e.d.}\label{sec:P6edprop}

\subsection{Squares of $P_6$-free graphs that have an e.d. are hole-free}\label{sec:G2holefree}

The main result of this subsection is Theorem~\ref{P6frG2holefree} which shows that for any $P_6$-free graph $G$ with an e.d., its square $G^2$ is hole-free.
This result is based on the unpublished thesis \cite{Friese2013}. It would imply that ED is solvable in polynomial time for $P_6$-free graphs if the MWIS problem for hole-free graphs is solvable in polynomial time, but the complexity of the MWIS problem for hole-free graphs is an open question. We will use Theorem~\ref{P6frG2holefree}, however, in subsequent sections for finding a polynomial time solution for ($P_6$, house)-free graphs (($P_6$, bull)-free graphs, respectively).

\begin{theorem}\label{P6frG2holefree}
If $G$ is a $P_6$-free graph that has an e.d., then $G^2$ is $C_k$-free for any $k \ge 5$.
\end{theorem}

For the proof of Theorem~\ref{P6frG2holefree}, we collect some subsequently described facts.
As in the proof of Theorem~\ref{P6HHDfrG2chordal}, let $D$ be an e.d. of $G$ and suppose to the contrary that $G^2$ contains an induced $C_k$ $C$, $k \ge 5$, with
$R(C)=\{v_1,\ldots,v_k\}$, $A(C)$ consisting of auxiliary vertices (if $d_G(v_i,v_{i+1})=2$ then the auxiliary vertex $x_i$ denotes a common neighbor of $v_i,v_{i+1}$ in $G$), and $V(C)=R(C) \cup A(C)$. Again, Observation~\ref{easyfacts} holds. We claim that $|A(C)| \ge k-1$:
\begin{lemma}\label{atmostonetype2}
For at most one $i \in \{1,\ldots,k\}$, $d_G(v_i,v_{i+1}) = 1$.
\end{lemma}

\noindent
{\bf Proof.}
Assume to the contrary that there are $i,j \in \{1,\ldots,k\}$, $i \neq j$, with $d_G(v_i,v_{i+1}) = 1$ and $d_G(v_j,v_{j+1}) = 1$. Without loss of generality, let $i=1$; then $2 < j \le k-1$.
If $d_G(v_2,v_j)=2$ and $d_G(v_{j+1},v_1)=2$, then $\{v_1,v_2,v_j,v_{j+1}\}$ induces a $C_4$ in $G^2$, which is impossible since $k \ge 5$. Thus, without loss of generality, let $d_G(v_2,v_j) > 2$ and let $P_{2,j}$ denote a shortest path in $C$ between $v_2$ and $v_j$ containing at least two vertices; but then, the induced subgraph $G[\{v_1,v_2,v_j,v_{j+1}\} \cup V(P_{2,j})]$ contains a $P_6$, which is a contradiction.
\qed

\medskip

If $d_G(v_i,v_{i+1})=1$ then we say that $v_i$ and $v_{i+1}$ are of {\em type $2$}, otherwise they are of {\em type $1$}.

\begin{lemma}\label{auxnonadj}
If $x_ix_j \notin E$, where $j \notin \{i-1, i+1\}$,
then $v_{i+1}v_j \notin E$, $v_{j+1}v_i \notin E$ and $x_i,x_j$ have a common neighbor $x_r$, $r \in \{i+1, i+2, \dots, j-1 \}$ and a common neighbor $x_s$, $s \in \{ j+1, j+2, \dots, i-1 \}$.
\end{lemma}

\noindent
{\bf Proof.}
Without loss of generality, assume that $x_1x_j \notin E$. Then, since $\{v_1,x_1,v_2$, $v_j,x_j,v_{j+1}\}$ does not induce a $P_6$ in $G$ and $k \ge 5$, we have $v_2v_j \notin E$ and similarly, $v_{j+1}v_1 \notin E$. Clearly, also $v_2v_{j+1} \notin E$ and $v_1v_j \notin E$ since $C$ is a $C_k$ in $G^2$, $k \ge 5$.
Now consider a shortest path between $x_1$ and $x_j$ in any arc of $C$ limited by these vertices - if it contains at least two vertices then there is a $P_6$. Thus, $x_1$ and $x_j$ have a common neighbor on both sides of the cycle which necessarily is an auxiliary vertex.
\qed

\begin{lemma}\label{notype1D}
If $v_i \in D$ then $v_i$ is not of type $1$.
\end{lemma}

\noindent
{\bf Proof.}
Assume without loss of generality that $v_1 \in D$ is of type 1. Then $v_k \notin D$ and $v_2 \notin D$; let $d_k,d_2 \in D$ with $d_kv_k \in E$ and $d_2v_2 \in E$. Obviously, $d_2 \neq d_k$ since $v_k$ and $v_2$ are nonadjacent in $G^2$. Now, if $x_1x_k \in E$ then $\{d_2,v_2,x_1,x_k,v_k,d_k\}$ induces a $P_6$, and if $x_1x_k \notin E$ then $\{d_2,v_2,x_1,v_1,x_k,v_k\}$ induces a $P_6$, which is a contradiction.
\qed

\begin{lemma}\label{V(C)Datmost1}
$|V(C) \cap D| \le 1$.
\end{lemma}

\noindent
{\bf Proof.}
Assume to the contrary that $|V(C) \cap D| \ge 2$. Note first that $|D \cap \{x_1,\ldots,x_k\}| \le 1$ since auxiliary vertices are either adjacent or share a common real neighbor $v_i$ or, by Lemma~\ref{auxnonadj}, share a common auxiliary neighbor $x_j$. Moreover, $|D \cap \{v_1,\ldots,v_k\}| \le 1$ since by Lemma~\ref{notype1D}, type 1 vertices are not in $D$ and by Lemma~\ref{atmostonetype2}, there are at most two (adjacent) type 2 vertices in $V(C)$.

Now assume without loss of generality that $v_1v_k \in E$ and $v_1 \in D$ as well as $x_i \in D$. Then obviously $i>1$, $v_{i+1}v_k \notin E$ and $x_1x_i \notin E$.
Thus, by Lemma~\ref{auxnonadj}, $x_1$ and $x_i$ have a common neighbor $w$ which is either the real vertex $v_2$ in the case $i=2$ (in which case $\{v_k,v_1,x_1,v_2,x_2,v_3\}$ induces a $P_6$), or an auxiliary vertex $x_j$ with $1<j<i$ but now $\{v_k,v_1,x_1,x_j,x_i,v_{i+1}\}$ induces a $P_6$, which is a contradiction.
\qed

\begin{lemma}\label{auxDadj}
If $x_i \in D$ then $x_ix_j \in E$ for all $j \notin \{i-1, i+1\}$.
\end{lemma}

\noindent
{\bf Proof.}
Without loss of generality, let $x_1 \in D$ and suppose to the contrary that there exists an $i \in \{3,\ldots,k-1\}$ with $x_1x_i \notin E$. We first claim:
\begin{clai}\label{commonneighb}
$v_i,x_i$ and $v_{i+1}$ are dominated by a common neighbor $d_i \in D \setminus V(C)$.
\end{clai}

\noindent
{\em Proof of Claim $\ref{commonneighb}$.} By Lemma~\ref{V(C)Datmost1} and the assumption that $x_1 \in D$, $v_i$ has to be dominated by a vertex $d_i \in D \setminus V(C)$. By Lemma~\ref{auxnonadj}, $x_1$ and $x_i$ have a common neighbor $x_j$, $j \in \{ i+1, i+2, \dots, k \}$. Since $\{d_i,v_i,x_i,x_j,x_1,v_2\}$ does not induce a $P_6$, we have $d_ix_i \in E$. Analogously, the $D$-vertex $d_{i+1}$ dominating $v_{i+1}$ must dominate $x_i$ as well, which by the e.d. property implies $d_i=d_{i+1}$ showing Claim~$\ref{commonneighb}$. $\diamond$

\medskip

By Lemma~\ref{atmostonetype2}, there is at least one auxiliary vertex $x_{i-1}$ or $x_{i+1}$, say without loss of generality, $x_{i+1}$ exists. Now, $x_1x_{i+1} \notin E$ since $\{v_2,x_1,x_{i+1},v_{i+1},d_i,v_i\}$ does not induce a $P_6$. Then there is a common neighbor $s$ of $x_1$ and $x_{i+1}$: Either $s=v_1$ or by Lemma~\ref{auxnonadj}, $s=x_j$ for some $j \in \{ i+2, i+3, \dots, k \}$. Now the induced subgraph 
$G[\{v_2,x_1,s,x_{i+1},v_{i+1},d_i,v_i\}]$ has at most one possible chord $d_ix_{i+1} \in E$ but in any case contains a $P_6$, which is a contradiction.
\qed

\begin{lemma}\label{noauxD}
$D \cap A(C) = \emptyset$
\end{lemma}

\noindent
{\bf Proof.}
Assume to the contrary that $D \cap A(C) \neq \emptyset$; without loss of generality, let $x_1 \in D$. By Lemma~\ref{auxDadj}, all auxiliary vertices $x_i$ with $3 \leq i < k$ are dominated by $x_1$. If there were two or more of them, say $x_i, x_j \in V(C)$ with $3 \leq i < j < k$ then either $x_ix_j \in E$ and $\{ d_i, v_i, x_i, x_j, v_{j+1}, d_{j+1} \}$ induces a $P_6$, or $x_i x_j \notin E$ and $\{ d_i, v_i, x_i, x_1, x_j, v_{j+1} \}$ induces a $P_6$ which are contradictions. Hence, there is at most one auxiliary vertex $x_i$ with $3 \leq i < k$. By Lemma \ref{atmostonetype2}, we conclude $k=5$ and either $d_G(v_3,v_4)=1$ or $d_G(v_4,v_5)=1$; without loss of generality, assume $v_3v_4 \in E$. Then $d_G(v_1,v_5) = d_G(v_4,v_5) = d_G(v_2,v_3) = 2$. By Lemma~\ref{auxDadj}, $x_1x_4 \in E$. Since $\{v_2,x_2,v_3,v_4,x_4,v_5\}$ does not induce a $P_6$, we have $x_2x_4 \in E$. Since $\{v_1,x_1,v_2,x_2,v_3,v_4\}$ does not induce a $P_6$, we have $x_1x_2 \in E$. Now $x_2$ is dominated by $x_1$ and thus, $\{d_3,v_3,x_2,x_4,v_5,d_5\}$ induces a $P_6$.
This final contradiction shows Lemma~\ref{noauxD}.
\qed

\begin{corollary}\label{nocommonDVvert}
No $D$-vertex can dominate two vertices of $\{v_1,\ldots,v_k\}$ such that at least one of them is of type $1$.
\end{corollary}

\noindent
{\bf Proof.}
Without loss of generality, assume that $d \in D$ dominates $v_1$ and $v_2$ and $v_2$ is a type 1 vertex, i.e., $x_1$ and $x_2$ exist. Then, $(V(C) \setminus \{x_1\}) \cup \{d\}$ realizes a $C_k$, $k \ge 5$, in $G^2$, where an auxiliary vertex is in $D$ which contradicts Lemma~\ref{noauxD}.
\qed

\begin{lemma}\label{nocommonDXvert}
If $v_i$ is a type $1$ vertex then the $D$-vertex $d_i$ dominating $v_i$ cannot dominate both of $x_{i-1},x_i$.
\end{lemma}

\noindent
{\bf Proof.}
Without loss of generality, assume that $d_1 \in D$ dominates $v_1,x_k,x_1$. Then there are $d_2,d_k \in D$ with $d_2v_2 \in E$ and $d_kv_k \in E$ and $d_2 \neq d_k$. By Corollary~\ref{nocommonDVvert}, $d_2 \neq d_1$ and $d_k \neq d_1$. Now there is a $P_6$ in the induced subgraph $G[\{d_2,v_2,x_1,v_1,x_k,v_k,d_k\}]$, which is a contradiction.
\qed

\begin{lemma}\label{nocommonDXdisjvert}
If for $x_i$ and $x_j$ with $j \notin \{i-1, i+1\}$, there are no edges between $v_i,v_{i+1}$ and $v_j,v_{j+1}$ then $x_i$ and $x_j$ have different $D$-neighbors.
\end{lemma}

\noindent
{\bf Proof.}
Assume that $d \in D$ dominates $x_i$ and $x_j$. Then by Lemma~\ref{noauxD}, $d \notin A(C)$, and by the distance assumptions of $C$, $d \notin R(C)$. Moreover, since $D$ is an e.d., $v_i,v_{i+1},v_j,v_{j+1} \notin D$.
By Corollary~\ref{nocommonDVvert}, $d$ misses $v_i$ or $v_{i+1}$, and $d$ misses $v_j$ or $v_{j+1}$; without loss of generality say, $d$ misses $v_i$
and $d$ misses $v_{j+1}$. By the same argument, for the $D$-neighbors $d_i$ of $v_i$ and $d_{j+1}$ of $v_{j+1}$, $d_i \neq d_{j+1}$. Now the induced subgraph  $G[\{d_i,v_i,x_i,d,x_j,v_{j+1},d_{j+1}\}]$ contains a $P_6$, which is a contradiction.
\qed

\begin{lemma}\label{notype2}
For all $i \in \{1,\ldots,k\}$, $v_i$ is of type $1$.
\end{lemma}

\noindent
{\bf Proof.}
Assume to the contrary that there are vertices of type 2 in $R(C)$: Without loss of generality, say $v_1$ and $v_k$ are of type 2, i.e., $v_1v_k \in E$. By Lemma~\ref{atmostonetype2}, we know that for $i \neq 1,k$, $v_i$ is of type 1. By Corollary~\ref{nocommonDVvert}, all such $v_i$ are dominated by personal $D$-neighbors.

We have $x_1x_{k-1} \in E$ since $\{v_{k-1},x_{k-1},v_k,v_1,x_1,v_2\}$ does not induce a $P_6$, and similarly, $x_{k-2}x_{k-1} \in E$ and $x_2x_1 \in E$.

Moreover, $d_2x_1 \in E$ or $d_2x_{k-1} \in E$ or $d_{k-1}x_1 \in E$ or $d_{k-1}x_{k-1} \in E$ since $\{d_2,v_2,x_1$, $x_{k-1},v_{k-1},d_{k-1}\}$ does not induce a $P_6$; without loss of generality, let
$x_{k-1}$ be dominated by $d_2$ or $d_{k-1}$. Since $\{d_{k-2},v_{k-2},x_{k-2}$, $x_{k-1},v_k,v_1\}$ does not induce a $P_6$, we have $d_{k-2}x_{k-2} \in E$. Then by Lemma~\ref{nocommonDXvert}, $d_{k-2}x_{k-3} \notin E$.

We claim that $x_1x_{k-2} \in E$: If $x_1x_{k-2} \notin E$ then by Lemma~\ref{auxnonadj}, $x_1$ and $x_{k-2}$ have a common neighbor $x_i$, $1 < i < k-2$ but then $\{v_{k-1},x_{k-2},x_i,x_1,v_1,v_k\}$ induces a $P_6$, which is a contradiction.
Moreover, we claim that $x_1x_{k-3} \in E$: For $k=5$, there is nothing to show. Now let $k>5$. If $x_1x_{k-3} \notin E$ then by Lemma~\ref{auxnonadj}, $x_1$ and $x_{k-3}$ have a common neighbor $x_i$, $1 < i < k-3$ but then $\{v_{k-2},x_{k-3},x_i,x_1,v_1,v_k\}$ induces a $P_6$, which is a contradiction.

Since $\{d_{k-2},v_{k-2},x_{k-3}$, $x_1,v_1,v_k\}$ does not induce a $P_6$, we have $d_{k-2}x_1 \in E$ which implies $d_2x_1 \notin E$, $d_{k-1}x_1 \notin E$ but now $\{d_{k-1},v_{k-1},x_{k-2}$, $x_1,v_1,v_k\}$ induces a $P_6$. This final contradiction shows Lemma~\ref{notype2}.
\qed

\medskip

Note that Lemmas~\ref{notype1D} and \ref{notype2} imply that $D \cap R(C) = \emptyset$.
The following summarizes the essential properties shown so far:

\begin{corollary}\label{summary}
For the cycle embedding $V(C)$, the following conditions hold:
\begin{enumerate}
\item[$(i)$] Every real vertex $v_i$ is of type $1$. Thus, $C$ consists of an alternating cycle of real and auxiliary vertices.
\item[$(ii)$] $V(C) \cap D = \emptyset$.
\item[$(iii)$] If a $D$-vertex dominates a real vertex then it dominates at most one of its auxiliary neighbors and no other real vertex.
\item[$(iv)$] If a $D$-vertex dominates an auxiliary vertex $x_i$ then it dominates another auxiliary vertex $x_j$ only if $j \notin \{i-1, i+1\}$.
\end{enumerate}
\end{corollary}

\begin{lemma}\label{realauxdom}
If a vertex $d \in D$ dominates a real vertex $v_i$ and an auxiliary vertex $x_j$, then $x_j$ is adjacent to a real vertex at distance at most $2$ of $v_i$.
\end{lemma}

\noindent
{\bf Proof.}
Assume not; i.e., $x_j$ misses $v_{i-1},v_i$ and $v_{i+1}$. Note that by this assumption, $j \notin \{i-2,i-1,i,i+1\}$ and thus, $v_{i-1},v_i,v_{i+1},v_j,v_{j+1}$ are five distinct vertices. We claim:
\begin{clai}\label{commonDneighbii+1}
$x_i$ and $v_{i+1}$ have a common $D$-neighbor $d_{i+1}$, and $x_{i-1}$ and $v_{i-1}$ have a common $D$-neighbor $d_{i-1}$.
\end{clai}

\noindent
{\em Proof of Claim $\ref{commonDneighbii+1}$.} By the previous lemmas, every $v_i$ has its personal $D$-neighbor $d_i$. We first show that $x_jx_{i-1} \in E$ and $x_jx_i \in E$. Assume to the contrary that $x_jx_{i-1} \notin E$. Then, since $G[\{d_{i-1},v_{i-1},x_{i-1},v_i,d,x_j,v_j\}]$ does not contain a $P_6$, we have both $dx_{i-1} \in E$ and $d_{i-1}x_{i-1} \in E$ which is impossible by the e.d. property. Thus, $x_{i-1}x_j \in E$ follows, and by symmetry, we also have $x_ix_j \in E$. Now, since $\{d_{i-1},v_{i-1},x_{i-1},x_j,v_j,d_j\}$ does not induce a $P_6$ and $\{d_{i-1},v_{i-1},x_{i-1},x_j,v_{j+1},d_{j+1}\}$ does not induce a $P_6$, we have either $d_{i-1}x_{i-1} \in E$ or both $d_jx_{i-1} \in E$ and $d_{j+1}x_{i-1} \in E$; the last is impossible by the e.d. property. Thus, $d_{i-1}x_{i-1} \in E$ follows. Again by symmetry, we have $d_{i+1}x_i \in E$ which shows Claim~$\ref{commonDneighbii+1}$. $\diamond$

\medskip

By Corollary~\ref{summary}, we have $d_{i+1}x_{i+1} \notin E$, $d_{i-1}x_{i-2} \notin E$, $d_{i+1}x_{i-2} \notin E$, and $d_{i-1}x_{i+1} \notin E$.
Since $\{d_{i+1},v_{i+1},x_{i+1},x_{i-2},v_{i-1},d_{i-1}\}$ does not induce a $P_6$, we have $x_{i+1}x_{i-2} \notin E$. By Lemma~\ref{auxnonadj}, there is a common neighbor of $x_{i-2}$ and $x_{i+1}$ in $\{x_{i-1},x_i\}$, say  $x_ix_{i+1} \in E$ and $x_ix_{i-2} \in E$ but now 
$\{v_{i+2},x_{i+1},x_i$, $x_{i-2},v_{i-1},d_{i-1}\}$ induces a $P_6$. This final contradiction shows Lemma~\ref{realauxdom}.
\qed

\medskip

\noindent
{\bf Proof of Theorem~\ref{P6frG2holefree}.}
Let $G$ be a $P_6$-free graph that has an e.d. $D$, and assume that there is a $C_k$ $C$, $k \ge 5$, in $G^2$ with a cycle embedding as described above.
Since $G[\{d_1,v_1,x_1$, $v_2,x_2,v_3,d_3\}]$ does not contain an induced $P_6$, one of the vertices $x_1,x_2$ is dominated by $d_1$ or $d_3$; without loss of generality, let $x_2$ be this vertex. Then $x_2d_2 \notin E$. Moreover, by
Lemma~\ref{realauxdom}, we have $x_4d_2 \notin E$. We claim that $x_2x_4 \notin E$: Assume $x_2x_4 \in E$. Then,
since $\{d_2,v_2,x_2,x_4,v_4,d_4\}$ does not induce a $P_6$, we obtain $x_4d_4 \in E$, and since $\{d_2,v_2,x_2,x_4,v_5,d_5\}$ does not induce a $P_6$, we obtain $x_4d_5 \in E$, which contradicts the e.d. property.

Thus, $x_2x_4 \notin E$ holds. Then by Lemma~\ref{auxnonadj}, $x_2$ and $x_4$ have the common neighbor $x_3$: $x_2x_3 \in E$ and $x_3x_4 \in E$.

Since $\{d_2,v_2,x_2,x_3,x_4,v_5\}$  does not induce a $P_6$, we obtain $d_2x_3 \in E$, which implies $x_3d_5 \notin E$. Moreover, by
Lemma~\ref{realauxdom}, $x_2d_5 \notin E$.

Since $\{v_2,x_2,x_3,x_4,v_5,d_5\}$  does not induce a $P_6$, we obtain $d_5x_4 \in E$. Moreover, we claim that $x_1x_4 \notin E$ holds:

If $x_1x_4 \in E$ then since $\{d_4,v_4,x_4,x_1,v_1,d_1\}$  does not induce a $P_6$, we obtain $x_1d_1 \in E$, and since $\{d_4,v_4,x_4,x_1,v_2,d_2\}$ 
 does not induce a $P_6$, we obtain $x_1d_2 \in E$, which contradicts the e.d. property. Thus, $x_1x_4 \notin E$.

By Lemma~\ref{auxnonadj}, $x_1$ and $x_4$ have the common neighbor $x_2$ or $x_3$, but we have already $x_2x_4 \notin E$; thus, the common neighbor of $x_1$ and $x_4$ is $x_3$.

Since $\{d_1,v_1,x_1,x_3,v_4,d_4\}$ does not induce a $P_6$, we obtain $d_1x_1 \in E$.

Again by Lemma~\ref{auxnonadj}, $x_1$ and $x_4$ have the common neighbor $x_j$, $4 < j \le k$.

Since $G[\{d_2,v_2,x_1,x_j,x_4,v_4,d_4\}]$ does not contain a $P_6$, we obtain $d_2x_j \in E$ and $d_4x_j \in E$, which contradicts the e.d. property. This final contradiction shows Theorem~\ref{P6frG2holefree}.
\qed

\subsection{Odd antiholes in squares of $P_6$-free graphs that have an e.d.}\label{oddantiholes}

Our main reason for considering odd antiholes in squares of $P_6$-free graphs with an e.d. is the famous Strong Perfect Graph Theorem \cite{ChuRobSeyTho2006} saying that a graph is perfect if and only if it is odd-hole-free and odd-antihole-free. If one were able to exclude odd antiholes in the squares of $P_6$-free graphs with an e.d., it would mean that $G^2$ is perfect and thus, ED would be solvable in polynomial time for $P_6$-free graphs. Some partial results in this direction are described subsequently.

Throughout this subsection, let $G=(V,E)$ be a $P_6$-free graph with an e.d. $D$, and let $G^2=(V,E^2)$.
Let $C$ be an odd antihole in $G^2$ with real vertices $R(C)$ and auxiliary vertices $A(C)$ as before. Since by 
Theorem~\ref{P6frG2holefree}, we know that $C_5$ is impossible in $G^2$, we can assume that $C$ is a $\overline{C_{2k+1}}$ for $k \ge 3$. Obviously, $|D \cap R(C)| \le 2$ since the distance between any two $D$-vertices is at least 3; $D$ is an independent vertex set in $G^2$, and the independence number of an odd antihole is 2. The main result of this section, namely Theorem~\ref{P6EDoddanti}, is based on \cite{Friese2013} and shows that no real vertex of an odd antihole $C$ is in $D$:

\begin{theorem}\label{P6EDoddanti}
If $G$ is a $P_6$-free graph that has an e.d. $D$ and $C$ is an odd antihole in $G^2$, then $|D \cap R(C)|=0$.
\end{theorem}

The proof of Theorem~\ref{P6EDoddanti} is based on some lemmas given subsequently.
Let $C$ be an odd antihole in $G^2$. We say that two vertices $a,b \in R(C)$ are {\em co-adjacent} in $C$ if they are nonadjacent in $G^2$.

\begin{lemma}\label{abcde}
Let $a,b,c,d,e$ be a sequence of consecutive co-adjacent real vertices in $R(C)$. If $ac \in E$ then $be \notin E$.
\end{lemma}

\noindent
{\bf Proof.}
Assume that $ac \in E$ and $be \in E$. Then by the assumption and by the distance properties of $C$, $ad \notin E$, $ae \notin E$, $ce \notin E$, and $bd \notin E$. Moreover, we have $ad \in E^2$ which implies $d_G(a,d)=2$ and we have $bd \in E^2$ which implies $d_G(b,d)=2$ (otherwise, there is a contradiction to co-adjacency of $c$ and $d$ (of $d$ and $e$, respectively). Let $p$ ($q$, respectively) be a common $G$-neighbor of $a$ and $d$ ($b$ and $d$, respectively). Since $a$ and $b$ are
co-adjacent in $C$, we have $p \neq q$. Moreover, $p$ and $q$ are distinct from any other vertex in $\{a,b,c,d,e\}$ since none of these is adjacent to $d$. Now, we consider the induced subgraph $G[\{c,a,p,d,q,b,e\}]$; the only possible chord is $pq \in E$, and thus, we obtain a $P_6$ in any case, which is a contradiction that shows Lemma~\ref{abcde}.
\qed

\medskip

Note that the proof of Lemma~\ref{abcde} does not require the existence of an e.d. in~$G$.

\begin{lemma}\label{abc12}
Let $a \in R(C) \cap D$ and let $b,c \in R(C)$ be two co-adjacent vertices in $C$ with $ab \in E^2$ and $ac \in E^2$. Then $d_G(a,b)=1 \Rightarrow d_G(a,c)=2$ and $d_G(a,b)=2 \Rightarrow d_G(a,c)=1$.
\end{lemma}

\noindent
{\bf Proof.}
By assumption, $d_G(a,b) \le 2$ and $d_G(a,c) \le 2$, and since $b,c \in R(C)$ are co-adjacent in $C$, $d_G(a,b)=2$ or $d_G(a,c)=2$ holds. Suppose that $d_G(a,b)=2$ and $d_G(a,c)=2$; let $p$ ($q$, respectively) be a common $G$-neighbor of $a$ and $b$ ($a$ and $c$, respectively). Since $b$ and $c$ are co-adjacent in $C$, we have $p \neq q$. Obviously, $b \notin D$ and $c \notin D$; let $\lambda \in D$ ($\mu \in D$, respectively) be the $D$-neighbor of $b$ ($c$, respectively). Obviously,
$\lambda \neq \mu$. Now consider the path $G[\{\lambda, b, p, a, q, c, \mu\}]$; the only possible chord is $pq \in E$, and thus, we obtain a $P_6$ in any case, which is a contradiction that shows Lemma~\ref{abc12}.
\qed

\begin{corollary}\label{abcde12}
Let $a,b,c,d,e$ be a sequence of consecutive co-adjacent real vertices in $R(C)$ with $c \in D$. Then $d_G(c,e)=1 \Rightarrow d_G(a,c)=2$ and $d_G(c,e)=2 \Rightarrow d_G(a,c)=1$.
\end{corollary}

\noindent
{\bf Proof.}
Let $R(C)$ be given by the sequence of consecutive real vertices $c,d,e,v_1$, $\ldots,v_k,a,b$. Since $C$ is an odd antihole, $k$ must be an even number. Repeatedly applying Lemma~\ref{abc12} gives $d_G(c,e)=1 \Rightarrow d_G(c,v_1)=2 \Rightarrow d_G(c,v_2)=1 \Rightarrow \ldots \Rightarrow d_G(c,v_k)=1 \Rightarrow d_G(a,c)=2$.  Similarly, if we begin with $d_G(c,e) = 2$. This shows Corollary~\ref{abcde12}.
\qed

\medskip

Next we show that at most one of $R(C)$ is in $D$:

\begin{lemma}\label{atmostoneD}
$|D \cap R(C)| \le 1$.
\end{lemma}

\noindent
{\bf Proof.}
Suppose that $|R(C) \cap D|=2$; let $b,c \in R(C) \cap D$. Then $b$ and $c$ are co-adjacent. Let $a,b,c,d,e,f \in V(C)$ induce a sequence of six consecutive real vertices in $C$ forming a $\overline{P_6}$ in $G^2$. We first claim:
\begin{equation}\label{ac1bd2}
d_G(a,c) = 1 \Leftrightarrow d_G(b,d) = 2.
\end{equation}

\noindent
{\em Proof of} (\ref{ac1bd2}):
By Lemma~\ref{abc12}, we have: $d_G(b,d) = 2 \Rightarrow d_G(b,e) = 1 \Rightarrow d_G(c,e) = 2$. Then Corollary~\ref{abcde12} implies $d_G(a,c) = 1$. Conversely, by
Lemma~\ref{abc12}, we have: $d_G(b,d) = 1 \Rightarrow d_G(b,e) = 2 \Rightarrow d_G(b,f) = 1 \Rightarrow d_G(c,f) = 2 \Rightarrow d_G(c,e) = 1$, and by Corollary~\ref{abcde12}, we have $d_G(a,c) = 2$. $\diamond$

\medskip

Without loss of generality, we can assume that $d_G(a,c) = 1$ and $d_G(b,d) = 2$. By Lemma~\ref{abc12}, this implies $d_G(b,e) = 1$. This, however, contradicts Lemma~\ref{abcde} since now $ac \in E$ and $be \in E$ holds. This shows Lemma \ref{atmostoneD}.
\qed

\medskip

The final step in the proof of Theorem~\ref{P6EDoddanti} is to show that none of the real vertices of $C$ are in~$D$.

\medskip

\noindent
{\bf Proof of Theorem~\ref{P6EDoddanti}.}
By Lemma~\ref{atmostoneD},
$|R(C) \cap D| \le 1$.
Suppose that $|R(C) \cap D|=1$ with $R(C) \cap D=\{c\}$; let $a,b,c,d,e \in R(C)$ be a sequence of consecutive co-adjacent real vertices in $C$ (and thus, $\{a,b,c,d,e\}$ induce a $\overline{P_5}$ in $G^2$). By Corollary~\ref{abcde12}, we can assume without loss of generality that $d_G(a,c) = 1$ and $d_G(c,e) = 2$. Let $p$ be a common neighbor of $c$ and $e$ and let $\lambda \in D$ be the $D$-neighbor of $e$ and $\mu \in D$ be the $D$-neighbor of $d$. Obviously, $c,\lambda$, and $\mu$ are pairwise distinct since $c$ and $d$ as well as $d$ and $e$ are co-adjacent in $C$. By Lemma~\ref{abcde}, $be \notin E$ since $ac \in E$ holds. Since $d_G(b,e)=2$, $b$ and $e$ have a common neighbor, say $q$. Let $f$ be the other real vertex in $C$ that is co-adjacent to $e$; by Lemma~\ref{abc12}, $cf \in E$ holds. Obviously, $q \neq p$ since $b$ and $c$ are co-adjacent, and $q \neq \mu$ since $\mu e \notin E$.  Further, $q \notin \{a,b,c,d,e,f\}$ since none of $a,b,c,d,e,f$ is adjacent to $b$ and $e$. We claim:
\begin{equation}\label{bclambdamu}
bc \notin E, \mbox{ } b \lambda \notin E, \mbox{ and } b \mu \notin E.
\end{equation}

\noindent
{\em Proof of} (\ref{bclambdamu}):
Obviously, by assumption $bc \notin E$ holds since $b$ and $c$ are co-adjacent in $C$. If $b \lambda \in E$ then $\{b,\lambda,e,p,c,f\}$ would induce a $P_6$ in $G$. Thus, $b \lambda \notin E$. For the case $b \mu \in E$, we consider the induced subgraph $G[\{d,\mu,b,q,e,p,c,f\}]$; by assumption, the e.d., and distance properties, the only possible chords are $pq \in E$, $\mu q \in E$ and $bd \in E$. Thus, the subgraph contains a $P_6$ in any case, which implies $b \mu \notin E$.
$\diamond$

\medskip

Since $bq \in E$, (\ref{bclambdamu}) implies $q \neq \lambda$. Let $\varphi \in D$ be the $D$-neighbor of $b$. By (\ref{bclambdamu}), $\varphi$ is distinct from $c$, $\lambda$, and $\mu$. Moreover, $\varphi q \in E$ since otherwise, the induced subgraph $G[\{\varphi,b,q,e,p,c,f\}]$ has at most one chord, namely $pq \in E$, and thus, there would be a $P_6$ in any case. Moreover, $bd \notin E$ since otherwise, the path $G[\{\lambda,e,q,b,d,\mu\}]$ is a $P_6$. Thus, there is a common neighbor $r$ of $b$ and $d$, which is distinct from all previously considered vertices since none of them is adjacent to $b$ and $d$. Obviously, $d_G(a,d)=2$ since otherwise $c$ and $d$ are not co-adjacent in $C$. Thus, there is a common neighbor $s$ of $a$ and $d$ which, by the same reason, is distinct from all previously considered vertices. 

Now, we show that $s \varphi \in E$: Note that $r \varphi \notin E$ since otherwise, the induced subgraph $G[\{\lambda,e,q,b,r,d,\mu\}]$ contains at most one chord, namely $qr \in E$, and thus, there would be a $P_6$ in any case.
Now the induced subgraph $G[\{\varphi,b,r,d,s,a,c\}]$ has at most two chords, namely $\varphi s \in E$ and $rs \in E$ and both are necessary since otherwise, the subgraph contains a $P_6$. In particular, $\varphi$ dominates $s$. Finally we consider the induced subgraph $G[\{\lambda,e,p,c,a,s,d,\mu\}]$. By the previous arguments, the only possible chords are $ea \in E$, $p a \in E$ and $p s \in E$. Thus, in any case, the subgraph contains a $P_6$, which is a contradiction that finally shows
Theorem~\ref{P6EDoddanti}.
\qed

\subsection{$C_4$ in squares of $P_6$-free graphs that have an e.d.}\label{C4G2}

Let $G=(V,E)$ be a $P_6$-free graph with an e.d. $D$, and let $G^2=(V,E^2)$ as defined above.
By Theorem~\ref{P6frG2holefree}, we know that the square of a $P_6$-free graph with an e.d. is $C_k$-free for any $k \ge 5$. For considering the ED problem on some subclasses of $P_6$-free graphs, it is useful to analyze how a $C_4$ in $G^2$ can be realized. In particular, Lemma~\ref{notype3} is helpful in various cases, and 
Lemmas \ref{type2.1} and \ref{type2.2} are used for solving ED on ($P_6$,house)-free graphs. 

As before, let $C$ be a $C_4$ in $G^2$ with real vertices $R(C)=\{v_1,v_2,v_3,v_4\}$ such that $v_iv_{i+1}$ are adjacent in $G^2$ (index arithmetic is modulo 4), and with auxiliary vertices $A(C)$. Let the auxiliary vertex $x_i$ be a common neighbor of $v_i$ and $v_{i+1}$; $x_i \in A(C)$ if and only if $v_iv_{i+1} \notin E$.

For this subsection, we assume that $D \cap R(C) = \emptyset$.  This assumption is motivated by Theorem~\ref{P6EDoddanti} which says for a $P_6$-free graph $G$ with an e.d. $D$, in an odd antihole $C$ of $G^2$, no real vertex of $C$ is in $D$; subsequently, we will consider a $C_4$ that is an induced subgraph of an odd antihole in $G^2$ where $G$ is a $P_6$-free graph with an e.d.
Let $d_i \in D$ denote the $D$-neighbor of $v_i$. Clearly, $v_i$ and $v_{i+2}$ have distinct $D$-neighbors for $i=1$ and $i=2$. There are the following types:

\medskip

\noindent
{\bf Type 1.} $R(C)$ is dominated by two $D$-vertices; say, $v_1,v_2$ are dominated by $d_1 \in D$, and $v_3,v_4$ are dominated by $d_3 \in D$.

\medskip

\noindent
{\bf Type 1.1} $v_1v_2 \notin E$, $v_3v_4 \notin E$.

\begin{lemma}\label{type1.1}
For any $C_4$ of type $1.1$, $v_2v_3 \in E$ and $v_1v_4 \in E$ holds.
\end{lemma}

\noindent
{\bf Proof.}
Suppose to the contrary that $v_2v_3 \notin E$. Then, since $\{v_2,d_1,v_1$, $v_4,d_3,v_3\}$ does not induce a $P_6$, we have $v_1v_4 \notin E$.
Since $\{v_1,d_1,x_2$, $v_3,d_3,v_4\}$ does not induce a $P_6$, we have $d_1x_2 \notin E$ and analogously we have $d_3x_2 \notin E$ but now $\{v_1,d_1,v_2,x_2,v_3,d_3\}$ induces a $P_6$, which is a contradiction. This shows Lemma~\ref{type1.1}.
\qed

\medskip

\noindent
{\bf Type 1.2} $v_1v_2 \in E$, $v_3v_4 \notin E$.

\begin{lemma}\label{type1.2}
For any $C_4$ of type $1.2$, $d_3x_2 \in E$ and $d_3x_4 \in E$ holds.
\end{lemma}

\noindent
{\bf Proof.}
Since $\{v_3,d_3,v_4,x_4,v_1,v_2\}$ does not induce a $P_6$, we have $d_3x_4 \in E$, and since $\{v_4,d_3,v_3$, $x_2,v_2,v_1\}$ does not induce a $P_6$, we have $d_3x_2 \in E$, which shows Lemma~\ref{type1.2}.
\qed

\medskip

\noindent
{\bf Type 1.3} $v_1v_2 \in E$, $v_3v_4 \in E$.

\medskip

Since $v_1v_2 \in E$ and $v_3v_4 \in E$, we have $v_2v_3 \notin E$ and $v_4v_1 \notin E$.

\begin{lemma}\label{type1.3}
For any $C_4$ of type $1.3$, we have: If neither $d_1$ nor $d_3$ dominates $x_2$ then $x_2x_4 \in E$ and either $d_1x_4 \in E$ or $d_3x_4 \in E$.
\end{lemma}

\noindent
{\bf Proof.}
Clearly, $x_2 \notin D$, and if neither $d_1$ nor $d_3$ dominates $x_2$, let $d \in D$, $d \neq d_1, d_3$, be a $D$-neighbor of $x_2$. Suppose to the contrary that $x_2x_4 \notin E$. Since $\{d_1,v_2,x_2,v_3,v_4,x_4\}$ does not induce a $P_6$, we have $d_1x_4 \in E$, and since $\{d_3,v_3,x_2,v_2,v_1,x_4\}$ does not induce a $P_6$, we have $d_3x_4 \in E$, which is contradiction. Thus, $x_2x_4 \in E$. Now, if neither $d_1x_4 \in E$ nor $d_3x_4 \in E$, then $x_4$ is dominated by some $d' \in D$, $d' \neq d_1, d' \neq d_3$, and now $\{d_1,v_2,x_2,x_4,v_4,d_3\}$ induces a $P_6$, which is a contradiction. Thus, $d_1x_4 \in E$ or $d_3x_4 \in E$, which shows Lemma~\ref{type1.3}.
\qed

\medskip

\noindent
{\bf Type 2.} $R(C)$ is dominated by three distinct $D$-vertices; say, $v_1,v_2$ are dominated by $d_1 \in D$, $v_3$ is dominated by $d_3 \in D$, and $v_4$ is dominated by $d_4 \in D$, $d_3 \neq d_4$.

\medskip

\noindent
{\bf Type 2.1} $v_1v_2 \notin E$:


\begin{lemma}\label{type2.1}
For any $C_4$ $C$ of type $2.1$, the following conditions hold:
\begin{itemize}
\item[$(i)$] $v_3v_4 \notin E$, $v_2v_3 \notin E$, and $v_1v_4 \notin E$. The auxiliary vertices $x_2,x_3,x_4$ are pairwise adjacent in $G$.
\item[$(ii)$] $d_1x_3 \in E$. Moreover, $d_1x_2 \notin E$ or $d_1x_4 \notin E$, and $d_1x_2 \notin E$ implies $d_3x_2 \in E$, while $d_1x_4 \notin E$ implies $d_4x_4 \in E$.
\end{itemize}
\end{lemma}

\noindent
{\bf Proof.}
Suppose to the contrary that $v_3v_4 \in E$. Then, by the distance properties, $v_2v_3 \notin E$ and $v_1v_4 \notin E$.
We claim that $d_3x_4 \notin E$ since otherwise, $\{v_3,d_3,x_4,v_1,d_1,v_2\}$ induces a $P_6$.
Now, since $\{d_3,v_3,v_4,x_4,v_1,d_1\}$ does not induce a $P_6$, we have $d_1x_4 \in E$, but then $\{d_3,v_3,v_4,x_4,d_1,v_2\}$ induces a $P_6$, which is a contradiction. Thus, $v_3v_4 \notin E$ holds.

\medskip

Suppose to the contrary that $d_1x_3 \notin E$. Then we claim that $v_2v_3 \notin E$ and $v_1v_4 \notin E$: Suppose to the contrary that $v_2v_3 \in E$. Since $\{d_4,v_4,x_3,v_3,v_2,d_1\}$ does not induce a $P_6$, we have $d_4x_3 \in E$, but now $\{d_4,x_3,v_3,v_2,d_1,v_1\}$ induces a $P_6$, which is a contradiction. Thus, $v_2v_3 \notin E$, and by symmetry, also $v_1v_4 \notin E$.

\medskip

\noindent
Since $\{v_1,d_1,v_2,x_2,v_3,x_3\}$ does not induce a $P_6$, we have $x_2x_3 \in E$ or $d_1x_2 \in E$.

\noindent
Since $\{v_2,d_1,v_1,x_4,v_4,x_3\}$ does not induce a $P_6$, we have $x_3x_4 \in E$ or $d_1x_4 \in E$.

On the other hand, if $d_1x_2 \in E$ and $d_1x_4 \in E$, then the induced subgraph $G[\{d_3,v_3,x_2$, $d_1,x_4,v_4,d_4\}]$ contains a $P_6$. Thus, without loss of generality, let $d_1x_2 \notin E$ which implies $x_2x_3 \in E$, but now $\{v_1,d_1,v_2,x_2,x_3,v_4\}$ induces a $P_6$, which is a contradiction. Thus, $d_1x_3 \in E$ holds.

\medskip

Suppose that $v_2v_3 \in E$. Then, since $\{d_3,v_3,v_2,d_1,v_1,v_4\}$ does not induce a $P_6$, we have $v_1v_4 \notin E$.
Since $\{v_3,v_2,d_1,v_1,x_4,v_4\}$ does not induce a $P_6$, we have $d_1x_4 \in E$, but now $\{d_3,v_3,v_2,d_1,x_4,v_4\}$ induces a $P_6$, which is a contradiction. Thus, $v_2v_3 \notin E$ and by symmetry, also $v_1v_4 \notin E$ holds. Again, the same argument as before shows that either $d_1x_2 \notin E$ or
$d_1x_4 \notin E$; without loss of generality let $d_1x_2 \notin E$.

\medskip

\noindent
{\bf Case 1.} $d_1x_4 \notin E$.

\medskip

\noindent
Since $\{d_4,v_4,x_4,v_1,d_1,v_2\}$ does not induce a $P_6$, we have $d_4x_4 \in E$.

\noindent
Since $\{d_3,v_3,x_3,v_4,x_4,v_1\}$ does not induce a $P_6$, we have $x_3x_4 \in E$.

\noindent
Since $\{d_3,v_3,x_2$, $v_2,d_1,v_1\}$ does not induce a $P_6$, we have $d_3x_2 \in E$.

\noindent
Since $\{v_3,x_2,v_2$, $d_1,v_1,x_4\}$ does not induce a $P_6$, we have $x_2x_4 \in E$.

\noindent
Since $\{d_4,v_4,x_3,d_1,v_2,x_2\}$ does not induce a $P_6$, we have $x_2x_3 \in E$.

\noindent
Thus, there is only one possible realization of $C$ in Case 1.

\medskip

\noindent
{\bf Case 2.} $d_1x_4 \in E$.

\medskip

Recall that for the same reason as above, $d_3x_2 \in E$ holds.

\noindent
Since $\{d_4,v_4,x_3$, $d_1,v_2,x_2\}$ does not induce a $P_6$, we have $x_2x_3 \in E$.

\noindent
Since $\{v_3,x_2,v_2,d_1,x_4,v_4\}$ does not induce a $P_6$, we have $x_2x_4 \in E$.

\noindent
Since $\{d_3,v_3,x_3,v_4,x_4,v_1\}$ does not induce a $P_6$, we have $x_3x_4 \in E$.

\noindent
Thus, there is only one possible realization of $C$ in Case 2.

\medskip

Thus, Lemma~\ref{type2.1} is shown.
\qed

\medskip

Note that $\{d_1,x,y,v_2,v_3\}$ induces a house in $G$ if $d_1x_2 \notin E$, and thus, in any case of type 2.1, $G$ contains a house.

\medskip

\noindent
{\bf Type 2.2} $v_1v_2 \in E$. Then by the distance properties, $v_2v_3 \notin E$ and $v_1v_4 \notin E$.

\begin{lemma}\label{type2.2}
For any $C_4$ $C$ of type $2.2$, the following conditions hold:
\begin{itemize}
\item[$( i)$] If $v_3v_4 \in E$ then $d_4x_2 \in E$, $d_3x_4 \in E$ and $x_2x_4 \in E$.
\item[$(ii)$] If $v_3v_4 \notin E$ then $x_2,x_3,x_4$ are pairwise adjacent in $G$, $d_3x_4 \in E$ or $d_3x_3 \in E$, and $d_4x_2 \in E$ or $d_4x_3 \in E$.
Moreover, $d_1x_2 \notin E$ or $d_1x_4 \notin E$, $d_3x_2 \notin E$ or $d_3x_3 \notin E$, and $d_4x_4 \notin E$ or $d_4x_3 \notin E$. At most one of $x_2,x_4$ is dominated by a vertex $d \in D$, $d \neq d_1,d_3,d_4$.
\end{itemize}
\end{lemma}

\noindent
{\bf Proof.}
$(i)$: Assume that $v_3v_4 \in E$. 

\noindent
Since $\{v_1,v_2,x_2,v_3,v_4,d_4\}$ does not induce a $P_6$, we have $d_4x_2 \in E$. 

\noindent
Since $\{v_2,v_1,x_4,v_4,v_3,d_3\}$ does not induce a $P_6$, we have $d_3x_4 \in E$. 

\noindent
Since $\{d_3,x_4,v_1,v_2,x_2,d_4\}$ does not induce a $P_6$, we have $x_2x_4 \in E$.

\medskip

\noindent
$(ii)$: Since $\{v_1,v_2,x_2,v_3,x_3,v_4\}$ does not induce a $P_6$, we have $x_2x_3 \in E$. 

\noindent
Since $\{v_2,v_1,x_4,v_4,x_3,v_3\}$ does not induce a $P_6$, we have $x_3x_4 \in E$. 

\noindent
Since $\{v_3,x_2,v_2,v_1,x_4,v_4\}$ does not induce a $P_6$, we have $x_2x_4 \in E$.

\noindent
Since $\{d_3,v_3,x_3,x_4,v_1,v_2\}$ does not induce a $P_6$, we have $d_3x_4 \in E$ or $d_3x_3 \in E$. 

\noindent
Since $\{d_4,v_4,x_3,x_2,v_2,v_1\}$ does not induce a $P_6$, we have $d_4x_2 \in E$ or $d_4x_3 \in E$.

\noindent
Since $\{d_3,v_3,x_2,x_4,v_4,d_4\}$ does not induce a $P_6$, we have $d_1x_2 \notin E$ or $d_1x_4 \notin E$.

\noindent
Since $\{d_1,v_2,x_2,x_3,v_4,d_4\}$ does not induce a $P_6$, we have $d_3x_2 \notin E$ or $d_3x_3 \notin E$.

\noindent
Since $\{d_1,v_1,x_4,x_3,v_3,d_3\}$ does not induce a $P_6$, we have $d_4x_4 \notin E$ or $d_4x_3 \notin E$.

\medskip

\noindent
{\bf Case 1.} $d_1x_4 \in E$.

\medskip

Then, since $\{d_3,v_3,x_3,x_4,v_1,v_2\}$ does not induce a $P_6$, we have $d_3x_3 \in E$, and since $\{d_4,v_4,x_3$, $x_2,v_2,v_1\}$ does not induce a $P_6$, we have $d_4x_2 \in E$.
Thus, there is only one possible realization of $C$ in Case 1.

\medskip

\noindent
{\bf Case 2.} $d_1x_4 \notin E$.

\medskip

\noindent
{\bf Case 2.1} $d_1x_3 \in E$.

\medskip

Then, by the previous arguments, $d_3x_4 \in E$ and $d_4x_2 \in E$. Thus, there is only one possible realization of $C$ in Case 2.1.

\medskip

\noindent
{\bf Case 2.2} $d_1x_3 \notin E$.

\medskip

In this case, at most one of $x_2,x_4$ is dominated  by a vertex $d \in D$, $d \neq d_1,d_3,d_4$ since otherwise $\{d_3,v_3,x_2,x_4,v_4,d_4\}$ induces a $P_6$.

\medskip

This shows Lemma~\ref{type2.2}.
\qed

\medskip

Note that in any case of type 2.2, $G$ contains a house.

\medskip

\noindent
{\bf Type 3.} $R(C)$ is dominated by four pairwise distinct $D$-vertices $d_1,d_2,d_3,d_4$.

\medskip

This type is excluded by the following:

\begin{lemma}\label{notype3}
For at least one pair $i,j \in \{1,2,3,4\}$, $i \neq j$, $d_i=d_j$ holds.
\end{lemma}

\noindent
{\bf Proof.} Assume to the contrary that $d_1,d_2,d_3,d_4$ are pairwise distinct.

\medskip

\noindent
{\bf Case 1.} Let $v_2v_3 \in E$ and $v_1v_4 \in E$. Then, by the distance conditions, $v_1v_2 \notin E$ and $v_3v_4 \notin E$. Thus, $C$ has two auxiliary vertices, say, $x_1$ (seeing $v_1$ and $v_2$) and $x_3$ (seeing $v_3$ and $v_4$).

\medskip

Since $\{d_1,v_1,x_1,v_2,v_3,d_3\}$ does not induce a $P_6$, we have $d_1x_1 \in E$ or $d_3x_1 \in E$. Then $\{d_2,v_2,x_1,v_1,v_4,d_4\}$ induces a $P_6$, which is a contradiction. Thus, Case~1 is excluded.

\medskip

\noindent
{\bf Case 2.} Let $v_1v_4 \in E$, but $v_1v_2 \notin E$, $v_2v_3 \notin E$ and $v_3v_4 \notin E$.

\medskip

\noindent
Since $\{v_4,v_1,x_1,v_2,x_2,v_3\}$ does not induce a $P_6$, we have $x_1x_2 \in E$.

\noindent
Since $\{v_1,v_4,x_3,v_3,x_2,v_2\}$ does not induce a $P_6$, we have $x_2x_3 \in E$.

\noindent
Since $\{v_2,x_1,v_1$, $v_4,x_3,v_3\}$ does not induce a $P_6$, we have $x_1x_3 \in E$.

\noindent
Since $\{d_1,v_1,v_4,x_3,x_2,v_2\}$ does not induce a $P_6$, we have $d_1x_2 \in E$ or $d_1x_3 \in E$.

\noindent
Since $\{d_2,v_2,x_2,x_3,v_4,v_1\}$ does not induce a $P_6$, we have $d_2x_2 \in E$ or $d_2x_3 \in E$.

\noindent
Thus, by the e.d. property, either $d_1x_2 \in E$ and $d_2x_3 \in E$ or $d_1x_3 \in E$ and $d_2x_2 \in E$.

\noindent
Since $\{d_3,v_3,x_2,x_1,v_1,v_4\}$ does not induce a $P_6$, we have $d_3x_1 \in E$ or $d_3x_2 \in E$.

\noindent
Since $\{d_4,v_4,v_1,x_1,x_2,v_3\}$ does not induce a $P_6$, we have $d_4x_1 \in E$ or $d_4x_2 \in E$.

\noindent
Now, by the e.d. property, if $d_1x_2 \in E$ and $d_2x_3 \in E$ then $d_3x_1 \in E$ and $d_4x_1 \in E$, which contradicts the e.d. property, and the same for
$d_1x_3 \in E$ and $d_2x_2 \in E$. Thus, Case 2 is excluded.

\medskip

\noindent
{\bf Case 3.} For all $i \in \{1,2,3,4\}$, $v_iv_{i+1} \notin E$ (index arithmetic is modulo 4).

\medskip

We first claim that for every $i \in \{1,2,3,4\}$, $d_ix_{i+1} \notin E$ or $d_ix_{i+2} \notin E$ as well as $d_ix_i \notin E$ or $d_ix_{i+3} \notin E$.
Suppose to the contrary that  $d_1x_2 \in E$ and $d_1x_3 \in E$. Then the induced subgraph $G[\{d_2,v_2,x_2,v_3,x_3,v_4,d_4\}]$ contains a $P_6$, which is a contradiction, and similarly for $d_1x_1 \in E$ and $d_1x_4 \in E$.

Moreover, we claim that for every $i \in \{1,2,3,4\}$, $d_ix_{i+1} \notin E$ or $d_ix_{i+3} \notin E$ as well as $d_ix_i \notin E$ or $d_ix_{i+2} \notin E$:
Suppose to the contrary that $d_1x_2 \in E$ and $d_1x_4 \in E$. Then if $x_2x_4 \in E$, $\{d_2,v_2,x_2,x_4,v_4,d_4\}$ induces a $P_6$, and if $x_2x_4 \notin E$, $\{d_2,v_2,x_2,d_1,x_4,v_4\}$ induces a $P_6$, which is a contradiction.

\medskip

Without loss of generality, let $d_1x_1 \notin E$.

\medskip

\noindent
{\bf Case 3.1} $d_1x_2 \notin E$.

\medskip

\noindent
Since $\{d_1,v_1,x_1,v_2,x_2,v_3\}$ does not induce a $P_6$, we have $x_1x_2 \in E$.

\noindent
Since $\{d_1,v_1,x_1,x_2,v_3,d_3\}$ does not induce a $P_6$, we have $d_3x_1 \in E$ or $d_3x_2 \in E$.

\medskip

\noindent
{\bf Case 3.1.1} $d_3x_1 \in E$. 

\medskip

Then $d_3x_4 \notin E$.

\medskip

\noindent
Since $\{v_3,d_3,x_1,v_1,x_4,v_4\}$ does not induce a $P_6$, we have $x_1x_4 \in E$.

\noindent
Since $\{v_3,d_3,x_1,x_4,v_4,d_4\}$ does not induce a $P_6$, we have $d_4x_4 \in E$, which implies $d_4x_3 \notin E$.

\noindent
Since $\{v_4,d_4,x_2,x_1,v_1,d_1\}$ does not induce a $P_6$, we have $d_4x_2 \notin E$.

\noindent
Since $\{v_2,x_2,v_3,x_3,v_4,d_4\}$ does not induce a $P_6$, we have $x_2x_3 \in E$.

\noindent
Since $\{v_1,x_1,x_2,x_3,v_4,d_4\}$ does not induce a $P_6$, we have $x_1x_3 \in E$.

\noindent
Since $\{d_1,v_1,x_1$, $x_3,v_4,d_4\}$ does not induce a $P_6$, we have $d_1x_3 \in E$, but now $\{d_2,v_2,x_1,x_3,v_4,d_4\}$ induces a $P_6$, which is a contradiction.

\medskip

\noindent
{\bf Case 3.1.2} $d_3x_1 \notin E$. 

\medskip
 
Then, since $d_3x_1 \in E$ or $d_3x_2 \in E$, we have  $d_3x_2 \in E$ which implies $d_3x_3 \notin E$ and $d_3x_4 \notin E$.

\medskip

\noindent
Since $\{d_3,v_3,x_3$, $v_4,x_4,v_1\}$ does not induce a $P_6$, we have $x_3x_4 \in E$.

\noindent
Since the induced subgraph $G[\{v_1,d_1,x_3$, $v_3,x_2,v_2,d_2\}]$ does not contain a $P_6$, we have $d_1x_3 \notin E$.

\noindent
Since $\{d_1,v_1,x_4$, $x_3,v_3,d_3\}$ does not induce a $P_6$, we have $d_1x_4 \in E$.

\noindent
Since $\{d_2,v_2,x_2$, $x_4,v_4,d_4\}$ does not induce a $P_6$, we have $x_2x_4 \notin E$, and

\noindent
since $\{d_1,v_1,x_1$, $x_3,v_3,d_3\}$ does not induce a $P_6$, we have $x_1x_3 \notin E$.

\noindent
Since $\{d_3,x_2,x_1$, $v_1,x_4,v_4\}$ does not induce a $P_6$, we have $x_1x_4 \in E$.

\noindent
Since $\{d_4,v_4,x_4$, $x_1,x_2,v_3\}$ does not induce a $P_6$, we have $x_1d_4 \in E$, but now $\{d_3,v_3,x_3,x_4,x_1,v_2\}$ induces a $P_6$, which is a contradiction.

\medskip

\noindent
{\bf Case 3.2} $d_1x_2 \in E$ (and for none of the $d_i$, $i=1,2,3,4$, the non-adjacencies are as in Case 3.1).

\medskip

Then $d_3x_2 \notin E$ and thus $d_3x_1 \in E$ (else Case 3.1 applies to $d_3$). Moreover, $d_1x_2 \in E$ implies $d_1x_3 \notin E$, and analogously, we have $d_3x_4 \notin E$. Recall that $d_2x_3 \notin E$ or $d_2x_4 \notin E$; without loss of generality, we can assume that $d_2x_4 \notin E$.

\medskip

\noindent
Since $\{d_2,v_2,x_1,v_1,x_4,v_4\}$ does not induce a $P_6$, we have $x_1x_4 \in E$.

\noindent
Since $\{d_2,v_2,x_1,x_4,v_4,d_4\}$ does not induce a $P_6$, we have $d_4x_4 \in E$, which implies $d_4x_3 \notin E$.

\noindent
Since $\{d_4,v_4,x_3,v_3,x_2,v_2\}$ does not induce a $P_6$, we have $x_2x_3 \in E$.

\noindent
Since $\{d_2,v_2,x_2,x_3,v_4,d_4\}$ does not induce a $P_6$, we have $d_2x_3 \in E$ (which implies $d_3x_3 \notin E$).

\noindent
Since $\{d_1,v_1,x_4,v_4,x_3,v_3\}$ does not induce a $P_6$, we have $x_3x_4 \in E$, but now $d_1,v_1,x_4,x_3,v_3,d_3$ is a $P_6$, which is a contradiction.
Thus, also Case 3 is excluded.

\medskip

This finally shows Lemma~\ref{notype3}.
\qed

\begin{corollary}\label{GP6housefrC4D2}
If $G$ is $(P_6$, house$)$-free graph that has an e.d. $D$ and $C$ is a $C_4$ in $G^2$ such that none of its real vertices is in $D$, then $R(C)$ is dominated by only two $D$-vertices.
\end{corollary}

\section{ED for ($P_6$, house)-free graphs and ($P_6$, bull)-free graphs in polynomial time}\label{EDP6housefr}

Throughout this section, let $G$ be a $P_6$-free graph that has an e.d. $D$.
The aim of this section is to show that for ($P_6$, house)-free graphs (for ($P_6$, bull)-free graphs, respectively), the ED problem is solvable in polynomial time. Independently, for ($P_6$, bull)-free graphs, ED was solved in polynomial time by T. Karthick \cite{Karth2015} using a different approach.

\begin{theorem}\label{GP6housefrG2oddanti}
If $G$ is a $(P_6$, house$)$-free graph that has an e.d., then $G^2$ is odd-antihole-free.
\end{theorem}

\noindent
{\bf Proof.}
Let $G$ be a ($P_6$, house)-free graph with an e.d. $D$. Suppose to the contrary that $G^2$ contains an odd antihole $H$ with real vertices $v_1,v_2,\ldots,v_{2k+1}$, $k \ge 3$, that are consecutively {\em co-adjacent} (i.e., nonadjacent in $G^2$). By Theorem~\ref{P6EDoddanti}, $D \cap \{v_1,v_2,\ldots,v_{2k+1}\}=\emptyset$ holds.
Clearly, the neighborhood of any vertex $d \in D$ in $H$ is a clique in $G^2$, and the clique cover number of $H$ in $G^2$ is 3. Thus, the number of $D$-vertices dominating $H$ is at least 3. Without loss of generality, let $d_1$ dominate $v_1$ and let $d_2$ dominate $v_2$. Since $v_1$ and $v_2$ are co-adjacent in $H$, $d_1 \neq d_2$ holds. If there is a vertex $d \in D$, $d \neq d_1, d \neq d_2$ dominating a vertex in $v_4,v_5,\ldots,v_{2k}$, then there is a $C_4$ in $H$ that is dominated by at least three $D$-vertices, which contradicts Corollary~\ref{GP6housefrC4D2}. Thus, assume that $d_1$ and $d_2$ dominate all of $v_4,v_5,\ldots,v_{2k}$, and without loss of generality, let $d$ dominate $v_3$. Then consider the $C_4$ $C$ induced by $\{v_2,v_3,v_5,v_6\}$. By assumption, $d_1$ and $d_2$ dominate $v_5$ and $v_6$ and since $v_5$ and $v_6$ are co-adjacent in $C$, the $D$-vertices dominating $v_5$ and $v_6$ are distinct.
Thus, $C$ has three distinct $D$-vertices which contradicts Corollary~\ref{GP6housefrC4D2}. This shows Theorem \ref{GP6housefrG2oddanti}.
\qed

\begin{theorem}\label{GP6bullfrG2oddanti}
If $G$ is a $(P_6$, bull$)$-free graph that has an e.d., then $G^2$ is odd-antihole-free.
\end{theorem}

\noindent
{\bf Proof.}
Let $G$ be a ($P_6$, bull)-free graph with an e.d. $D$, and suppose to the contrary that $G^2$ contains an odd antihole $H$ with real vertices $v_1,v_2,\ldots,v_{2k+1}$, $k \ge 3$, that are consecutively co-adjacent. As before, $D \cap \{v_1,v_2,\ldots,v_{2k+1}\}=\emptyset$ holds, by Theorem~\ref{P6EDoddanti}. We first show:

\begin{clai}\label{C4twoedges}
For every $C_4$ in $H$ with real vertices $u_1,u_2,u_3,u_4$, for exactly two values of $i$, $1 \le i \le 4$, $u_iu_{i+1} \in E$ holds.
\end{clai}

\noindent
{\em Proof of Claim $\ref{C4twoedges}$.}
Let $u_1,u_2,u_3,u_4$ be a $C_4$ in $G^2$ with $d_G(u_i,u_{i+1}) \le 2$ (as before, let $x_i$ be a common neighbor of $u_i$ and $u_{i+1}$ if $u_iu_{i+1} \notin E$) and suppose that for at most one $i$, $u_iu_{i+1} \in E$ holds.


First suppose that there is exactly one edge $u_iu_{i+1}$, say $u_1u_2 \in E$. Then $u_2u_3 \notin E$, $u_3u_4 \notin E$, and $u_4u_1 \notin E$. Since $\{u_1,u_2,x_2,u_3,x_3,u_4\}$ does not induce a $P_6$, we have $x_2x_3 \in E$ but, now $\{u_2,x_2,u_3,x_3,u_4\}$ induces a bull, which is a contradiction.


Thus, for all $i$, $u_iu_{i+1} \notin E$ holds. We have seen already that if $x_ix_{i+1} \in E$ for some $i$, then $\{u_i,x_i,u_{i+1},u_{i+2},x_{i+1}\}$ induces a bull. Thus, for all $i$, $x_ix_{i+1} \notin E$. Since $\{u_1,x_1,u_2,x_2,u_3,x_3\}$ does not induce a $P_6$, we have $x_1x_3 \in E$ and similarly we have $x_2x_4 \in E$. By Lemmas \ref{type1.1} and \ref{notype3}, we know that the $C_4$ has exactly three $D$-vertices $d_1,d_3,d_4$; say $d_1$ sees $u_1$ and $u_2$, $d_3$ sees $u_3$, and $d_4$ sees $u_4$. Recall that $d_1x_2 \notin E$ and $d_1x_4 \notin E$ since $G$ is assumed to be bull-free.

\noindent
Since $\{u_1,d_1,u_2,x_2,u_3,d_3\}$ does not induce a $P_6$, we have $x_2d_3 \in E$.

\noindent
Since $\{u_2,d_1,u_1,x_4,u_4,d_4\}$ does not induce a $P_6$, we have $x_4d_4 \in E$.

\noindent
Since $\{u_2,x_2,u_3,x_3,d_3\}$ does not induce a bull, we have $x_3d_3 \in E$.

\noindent
Since $\{u_1,x_4,u_4,x_3,d_4\}$ does not induce a bull, we have $x_3d_4 \in E$, which is a contradiction showing Claim~$\ref{C4twoedges}$.
$\diamond$

\medskip

By Claim $\ref{C4twoedges}$, every $C_4$ in the odd antihole $H$ of $G^2$ has exactly two edges in $E$. We apply this as follows:
\begin{clai}\label{C4oddahalternating}
For all $i, 1 \le i \le 2k+1$, we have: If $v_iv_{i+2} \in E$ then $v_{i+1}v_{i+3} \in E$ $($index arithmetic is modulo $2k+1)$. In particular,
if for some $i$, $v_iv_{i+2} \in E$ then for all $i$, $1 \le i \le 2k+1$, $v_iv_{i+2} \in E$.
\end{clai}

\noindent
{\em Proof of Claim $\ref{C4oddahalternating}$.}
Let $v_1v_3 \in E$. Then, by the distance conditions, $v_1v_4 \notin E$ and $v_3v_{2k+1} \notin E$. Considering the $C_4$ in $G^2$ induced by $\{v_1,v_3,v_4,v_{2k+1}\}$, we have $v_4v_{2k+1} \in E$, which implies $v_5v_{2k+1} \notin E$. Considering the $C_4$ in $G^2$ induced by $\{v_1,v_4,v_5,v_{2k+1}\}$, we have $v_5v_1 \in E$, which implies $v_5v_2 \notin E$.  Considering the $C_4$ in $G^2$ induced by $\{v_1,v_2,v_4,v_5\}$, we have $v_2v_4 \in E$. Applying this repeatedly along the odd antihole $H$, we obtain $v_iv_{i+2} \in E$ for all $i$.
$\diamond$

\medskip

Now first assume that for one $i$, $v_iv_{i+2} \in E$ holds; say, $v_1v_3 \in E$.
Then we consider the $C_4$s with $v_1,v_{2k+1}$ and the opposite pairs $v_{i},v_{i+1}$, $3 \le i \le 2k-2$, and we obtain an alternating sequence of edges and non-edges for $v_1$, i.e, $v_1v_i \in E$ for all odd $i$, $3 \le i \le 2k-1$ and $v_1v_i \notin E$ for all even $i$, $4 \le i \le 2k-2$, and considering the $C_4$  in $G^2$ induced by $\{v_1,v_{2k-2},v_{2k-1},v_{2k+1}\}$, we obtain $v_{2k+1}v_{2k-1} \notin E$, which contradicts Claim~\ref{C4oddahalternating}.

\medskip

Thus, suppose that for all $i, 1 \le i \le 2k+1$, $v_iv_{i+2} \notin E$ holds.
Since by assumption, every $C_4$ has exactly two $E$-edges, we can assume that $v_1v_i \in E$ for some $i$. Then, by the distance conditions, $v_1v_{i-1} \notin E$, $v_1v_{i+1} \notin E$, $v_iv_2 \notin E$, and $v_iv_{2k+1} \notin E$. By the $C_4$ argument, $v_2v_{i-1} \in E$ and $v_{2k+1}v_{i+1} \in E$ follows. Repeatedly applying the distance argument and the $C_4$ argument implies that finally, for some $j$, $v_jv_{j+2} \in E$, which is a contradiction
that concludes the proof of Theorem~\ref{GP6bullfrG2oddanti}.
\qed

\begin{corollary}\label{GP6housebullfrEDpol}
For $(P_6$, house$)$-free graphs and $(P_6$, bull$)$-free graphs, the WED problem is solvable in polynomial time.
\end{corollary}

\noindent
{\bf Proof.}
First suppose that $G$ is ($P_6$, house)-free.
By Theorem~\ref{P6frG2holefree}, for a $P_6$-free graph $G$ with an e.d., $G^2$ is hole-free. By Lemma~\ref{GP6housefrG2oddanti}, $G^2$ is odd-antihole-free.
If $G^2$ is odd-hole-free and odd-antihole-free then, by the Strong Perfect Graph Theorem \cite{ChuRobSeyTho2006}, $G^2$ is perfect. By \cite{GroLovSch1981}, MWIS is solvable in polynomial time for perfect graphs. By Lemma~\ref{mainequived}, the ED problem on $G$ can be transformed into the MWIS problem on $G^2$. Thus, ED is solvable in polynomial time on 
($P_6$, house)-free graphs. By \cite{BraFicLeiMil2013}, the WED problem can be solved in polynomial time for the same class.

Now suppose that $G$ is ($P_6$, bull)-free. By Lemma~\ref{GP6bullfrG2oddanti}, $G^2$ is odd-antihole-free, and thus, $G^2$ is perfect. Hence, WED is solvable in polynomial time on ($P_6$, bull)-free graphs by the same arguments as above.
\qed

\section{Conclusion}

The main results of this paper are the following:

\begin{enumerate}
\item[$(i)$] Theorem~\ref{P6HHDfrG2chordal}: If $G$ is ($P_6$, HHD)-free and has an e.d., then $G^2$ is chordal, and thus, ED/WED is solvable in polynomial time for this class of graphs. This gives a dichotomy result for $P_k$-free chordal graphs since ED is \NP-complete for $P_7$-free chordal graphs.

\item[$(ii)$] Theorem~\ref{P6frG2holefree}: If $G$ is $P_6$-free and has an e.d., then $G^2$ is hole-free. This does not yet imply that ED for $P_6$-free graphs is solvable in polynomial time since the MWIS problem for hole-free graphs is an open question, but it has some applications for subclasses of $P_6$-free graphs.

\item[$(iii)$] Theorem~\ref{P6EDoddanti}: If $G$ is $P_6$-free and has an e.d., then odd antiholes in $G^2$ would have very special structure.

\item[$(iv)$] Theorems~\ref{GP6housefrG2oddanti} and \ref{GP6bullfrG2oddanti}: Using results on the structure of $C_4$ realizations in $G^2$, we obtain a polynomial time solution of ED/WED for ($P_6$, house)-free graphs and for ($P_6$, bull)-free graphs since in these cases, $G^2$ is perfect if $G$ has an e.d.
\end{enumerate}

For some other subclasses of $P_6$-free graphs, ED has been solved in polynomial time, such as for ($P_6$, $S_{1,2,2}$)-free graphs \cite{BraMilNev2013} and for ($P_6$, $S_{1,1,3}$)-free graphs \cite{Karth2015}. The complexity of ED for $P_6$-free graphs remains a challenging open question.
The following conjecture appeared in \cite{Friese2013}:

\medskip

\noindent
{\bf Conjecture.} If $G$ is a $P_6$-free graph that has an e.d., then $G^2$ is perfect.

\medskip

\noindent
{\bf Acknowledgements.} The first and second authors gratefully acknowledge support from the West Virginia University ADVANCE Sponsorship Program, and the first author thanks Van Bang Le for discussions about the Efficient Domination problem.

\begin{footnotesize}

\end{footnotesize}

\end{document}